\documentclass[preprint,12pt]{elsarticle}

\usepackage{amssymb}
\usepackage{amsmath}
\usepackage{wasysym}
\usepackage{graphicx}
\usepackage{dcolumn}
\usepackage[utf8]{inputenc}
\usepackage{xr}
\usepackage[T1]{fontenc}
\usepackage{mwe}
\usepackage[dvipsnames]{xcolor}
\usepackage{hyperref}
\usepackage{ulem}
\usepackage{stix}
\journal{Journal of Crystal Growth}

\begin{document}

\begin{frontmatter}

\title{Engineering 2D Surface Patterns with the VicCa Model}
\author{Marta A. Chabowska}
\ead{galicka@ifpan.edu.pl}
\author{Magdalena A. Za{\l}uska-Kotur} 

\affiliation{organization={Institute of Physics, Polish Academy of Sciences},
            addressline={Al. Lotnikow 32/46},
            city={Warsaw},
            country={Poland}}

\begin{abstract}
We employed the VicCA model to investigate the influence of step-edge potential on nucleation and pattern formation, aiming to gain deeper insights into island formation and growth. Our study explores fractal structures governed by general cellular automaton (CA) rules, as well as compact structures shaped by density-dependent attachment mechanisms. We demonstrate that modifications to the CA framework have a significant impact on surface patterning, emphasizing the critical role of adatom attachment rules and the substantial effect of potential well depth on the resulting surface morphology.
\end{abstract}







\end{frontmatter}


\section{Introduction}
Nucleation is a fundamental step in crystal growth, where atoms, ions, or molecules come together to form a stable cluster. Depending on the conditions, it can either facilitate or hinder the formation of the desired crystal structure.
In some cases, unwanted nucleation disrupts the smooth growth of layers, leading to rough surfaces or defects. Mastering this process is essential for achieving high-quality thin films, especially in semiconductor manufacturing. Conversely, controlled nucleation enables the creation of novel nanostructures, such as quantum dots, nanowires, and patterned surfaces, with applications in electronics, catalysis, and materials science.
Beyond technological applications, nucleation also plays a vital role in biological systems, such as protein crystallization and the formation of coral structures.
A deep understanding of island formation and growth is essential for precise control and design of surface structures.

To date, a considerable number of publications have been dedicated to the analysis of the formation of islands and their shapes. Part of the papers in the field investigate the directional solidification (DS) of monocrystals and alloys, which very often results in dendritic growth \cite{Chen, Mendoza, Kammer, Zhang}. In their analysis, Zhang et al. \cite{Zhang} employed a cellular automaton-finite difference (CA-FD) model, which was based on heat transfer and solute diffusion. They also proposed a modified shaped function model to predict the evolution behavior of a large number of dendrites during DS process. The transition from dendritic to compact shapes has often been analyzed using the Monte Carlo method or diffusion limited aggregation (DLA) \cite{Xiao-88,Bales,DeVita,Michely,Wang}. Xiao et al. in their work \cite{Xiao-88} simulated the morphological evolution of crystals growing from a vapor phase on a triangular lattice. Depending on various parameters used, the crystals were found to grow either fully compact faceted or with dendritic protrusions. In  Ref~\cite{Bales}  a linear stability analysis was used to examine the morphological transition from compact to fractal islands during epitaxial growth. The authors demonstrate that at low temperatures, the islands exhibit ramified structures, while at high temperatures, they adopt a compact structure. According to the authors, at intermediate temperatures, the islands assume a structure that is between the two extremes.  In their work, DeVita et al.~\cite{DeVita}, develop a method for simulating epitaxial growth, based on coarse graining the evolution equations of the probability functions for Kinetic Monte Carlo. As Bales et al. \cite{Bales}, they also demonstrate that at early stages of growth, at E/kT = 3.5, they obtain compact islands, while at zero temperature, they obtain dendritic islands. In the book~\cite{Michely} T. Michely and J. Krug present a comprehensive study of island shapes.
For instance, in the case of Au grown on Ru(0001), highly ramified, isotropic fractal islands form at room temperature. In contrast, the equilibrium shape of Pt islands grown on Pt(111) at high temperatures (700 K and above) is a hexagon with threefold symmetry. At lower temperatures (200 K), fractal-dendritic islands emerge. The authors note that such island formation is a common feature of low-temperature homoepitaxial growth on fcc(111) metal surfaces, including Pt(111), Ag(111), Ir(111), and Al(111). At intermediate temperatures, Pt islands take on a compact, triangular shape. The transition to fully compact islands, the authors suggest, occurs only when adatom transport around step corners becomes feasible.
Once this kinetic channel opens, step adatoms can respond to the binding energy difference between the two types of island edges, influencing the formation of triangular shapes. At low temperatures, Wang et al. \cite{Wang} found that Si islands adopt compact and irregular shapes. However, after the addition of Ag atoms, these accumulate at the step edges above T=300 K , leading to dendritic growth.
In their review, Yang et al. \cite{Yang-AFM} discussed the influence of interfacial instability on crystal morphology. When interfacial instability is absent or minimal, crystal growth follows a stable facet growth mode, resulting in polyhedral crystal shapes. With an intermediate degree of instability, hopper-shaped crystals emerge, while a further increase in instability leads to dendritic structures.
An interesting perspective on surface and island shape analysis can be found in the work of B. Klepka et al. \cite{Klepka}. Authors investigated the impact of pre-crystallisation states and the presence of polyanionic protein AGARP on the morphology of the emerging calcium carbonate (CaCO$_3$) phases in different conditions. The results demonstrated that  CaCO$_3$ phases grown with protein AGARP in the presence of carbon dioxide without and with 20\% of additional polymer exhibit smoother and rounded edges. In contrast, the CaCO$_3$ phases grown in the absence of protein AGARP exhibited sharper edges and a dendritic morphology.

Different surface patterns arise from the interplay between atomic dynamics and interatomic interactions. These interactions determine how adatoms move and where they attach to the crystal, shaping island growth and the final surface structure. The interaction between adatoms and bulk surface atoms can also be represented as a surface potential landscape that governs adatom motion. This perspective - describing surface dynamics in terms of motion within an energy potential - is increasingly used in crystal growth studies. Local potential maps, derived from density functional theory (DFT), have been published for various systems \cite{Michely,Akiyama-JJAP}. Our study focuses on understanding surface dynamics through the surface potential energy landscape and its crucial role in pattern formation.

We analyze surface dynamics using the (2+1) dimensional cellular automaton model (VicCA), which integrates CA attachment rules for crystal growth with Monte Carlo (MC) diffusional jumps in the adatom layer. This approach distinguishes different aspects of surface evolution, simplifying them into fundamental rules that govern island nucleation and growth.
Our study shows that even small modifications to CA attachment rules lead to significantly different surface patterns.  We demonstrate further that for a given set of CA rules, variations in the surface potential energy landscape significantly influence pattern formation, highlighting the critical role of step-edge potential in island growth.
We examine both fractal structures shaped by general CA rules and a broader class of compact structures, formed through density-dependent attachment to step rules. The shift from fractal to compact growth arises as CA attachment rules become more restrictive. However, the final morphology - defined by fractal dimensionality and the dendritic, square, or circular shapes of compact islands - is predominantly governed by potential depth, which regulates adatom density along the steps.

\section{The model}
Our (2~+~1)D vicinal Cellular Automaton model, simulates the evolution of a crystal surface and the diffusion of adatoms on its surface. Previous studies have explored this model in (1~+~1)D \cite{Krasteva-AIP,Krzyzew-JCG,Toktarbaiuly-prb,Krzyzew-CGD,Popova-CGD,Popova-CGD23} and (2~+~1)D \cite{MZK-crystals,Chabowska-ACS,Chabowska-Vac,Chabowska-PRB} context. The model combines a Cellular Automaton module for surface growth and a Monte Carlo module for adatoms diffusion. The CA module updates the surface in parallel based on pre-defined rules, while the MC module simulates adatoms diffusion sequentially.

Within this framework, each simulation time step encompasses three key processes: adatom diffusion across the surface, surface growth updates within the CA module, and the replenishment of adatoms to maintain their initial concentration, $c_0$. A single diffusion step is completed once every adatom has been visited, on average, once. This structured approach enables efficient large-scale simulations. Between two consecutive growth updates, each adatom attempts $n_{DS}$ diffusion jumps, though only those directed toward unoccupied neighboring lattice sites are successfully executed.

The model is based on a square lattice that defines the system, consisting of two distinct components: the crystal surface and the adatom layer. To investigate the initial stages of island formation, this work considers an initially flat surface. Additionally, a single atom is placed at the center of the terrace, serving as a growth seed. Periodic boundary conditions are applied. The incorporation of adatoms into the crystal structure follows the CA rules.

A crucial aspect of the model is the diffusion of adatoms, which drives the formation of surface patterns. This diffusion process is influenced by the energy potential landscape experienced by the atoms. We adopt the shape of this landscape based on findings from previous studies~\cite{Chabowska-Vac,Chabowska-PRB,Ohka-CGD,Akiyama-JCG20}, assuming the existence of a potential well at the bottom of the step. The depth of this potential well remains constant throughout the simulation and is given by $\beta E_V$, where $E_V$ represents the energy associated with the well, and $\beta=1/k_B T$, with $T$ denoting temperature and $k_B$ being the Boltzmann constant.

In the model, each particle diffuses independently. All jumps along terraces - except those near the step - occur with equal probability, which is set to 1 after a uniform selection of jump direction. The probability of a jump out of the well is given by $P = e^{-\beta E_V}$. The presence of the potential well plays a crucial role in localizing particles, increasing their density at the bottom of the step. In the following examples, we also examine the effect of a potential hill instead of a well, where the potential at the bottom of the step, $E_B$ , is higher rather than lower than at other sites. In this scenario, the probability of a jump into this site is given by $P = e^{-\beta E_B}$ , while the probability of a jump from this site is 1. As a result, the particle density at the step is lower than that on the terrace.

\section{Results and discussion}
The vicCA model facilitates crystal growth, leading to a diverse range of surface patterns, including meanders, bunches, pyramids, and nanowires. A key feature of the model is the spontaneous formation of islands on the surface. By implementing simple attachment rules for adatoms, the model effectively captures the complexity and variety of these structures. Since this study focuses on the formation and evolution of islands on the crystal surface, the following discussion will primarily explore their early stages of development.
\begin{figure}[hbt]
 \centering
a)\includegraphics[width=0.45\textwidth]{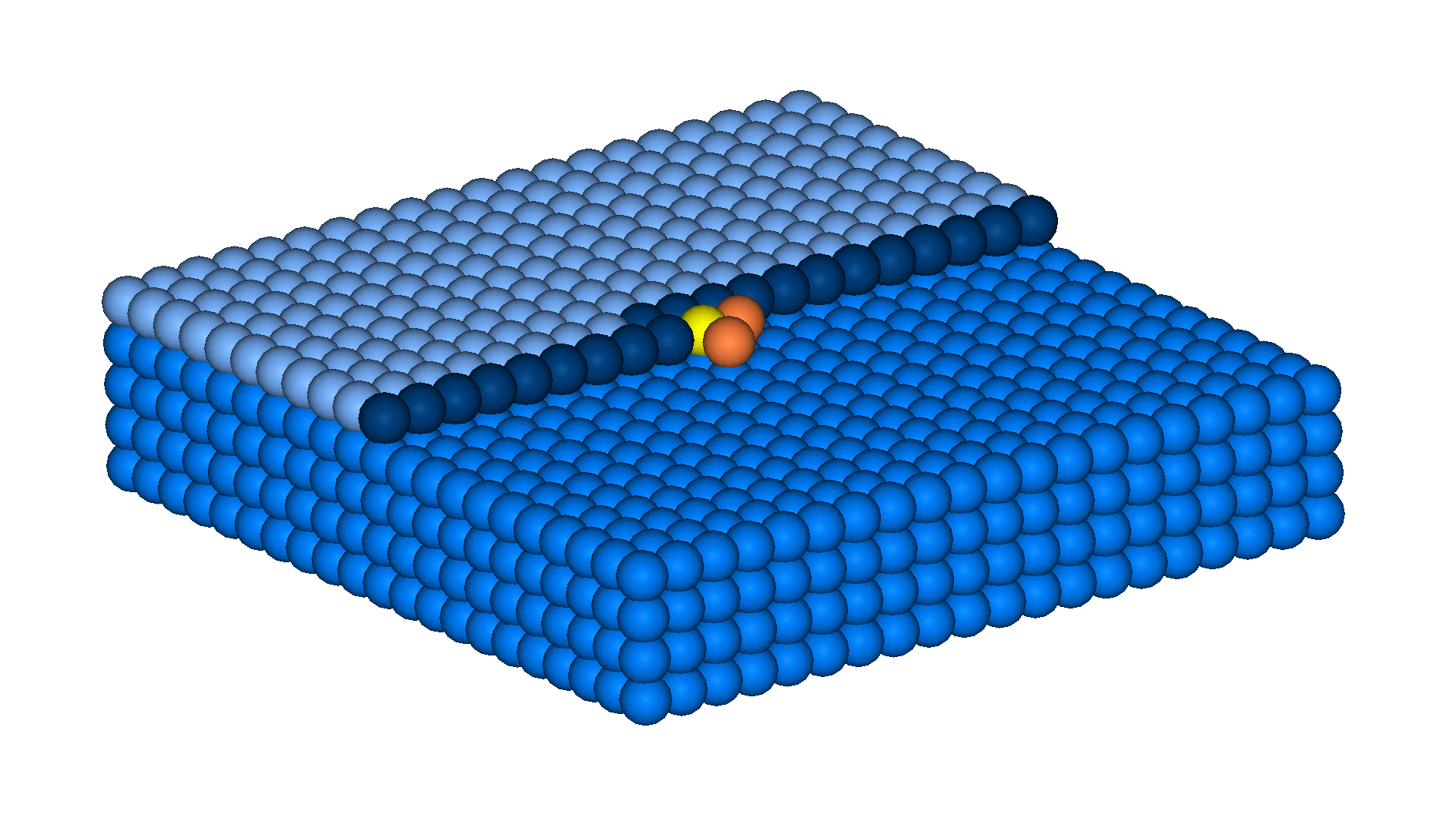}
b)\includegraphics[width=0.45\textwidth]{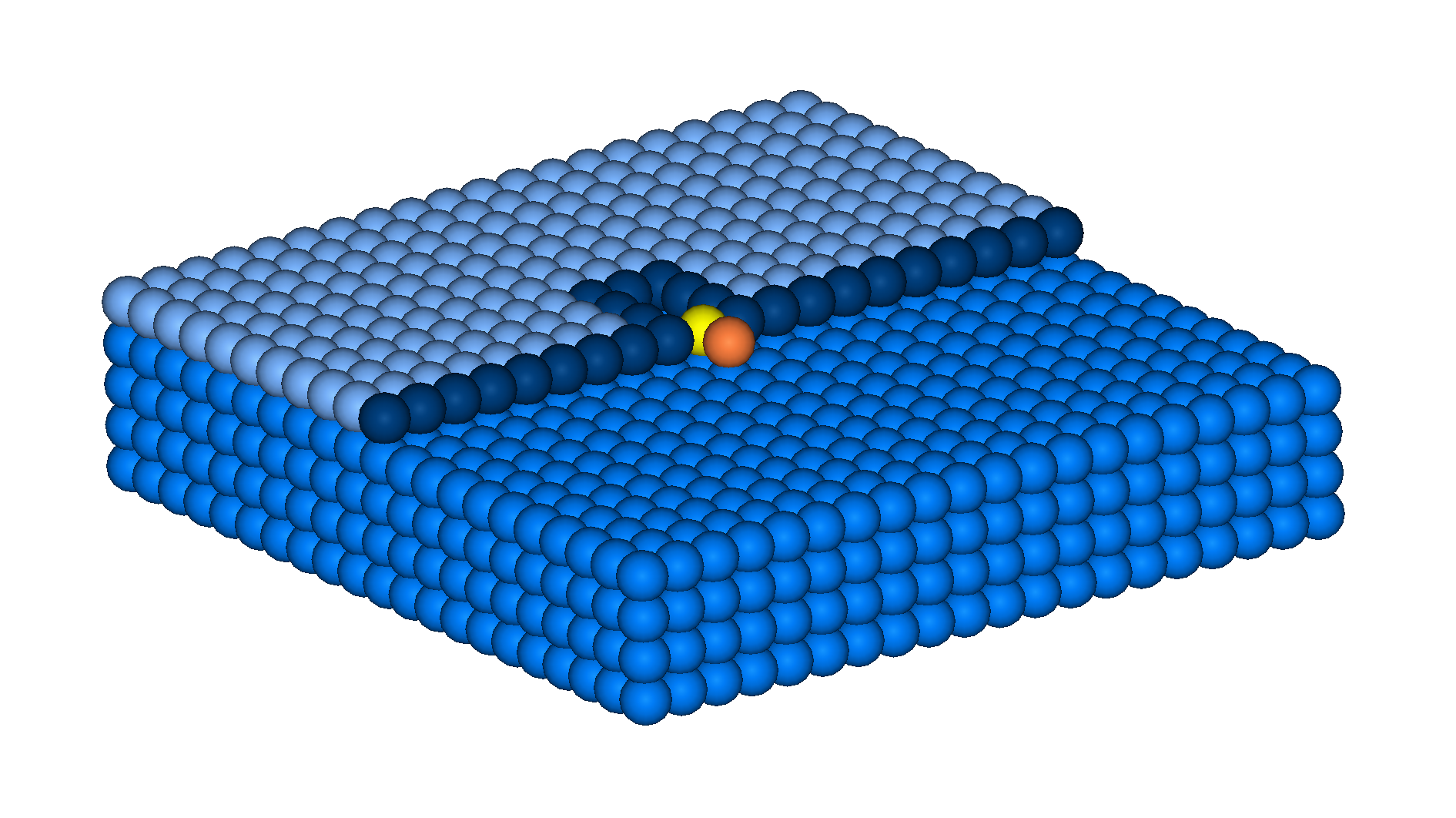}
c)\includegraphics[width=0.45\textwidth]{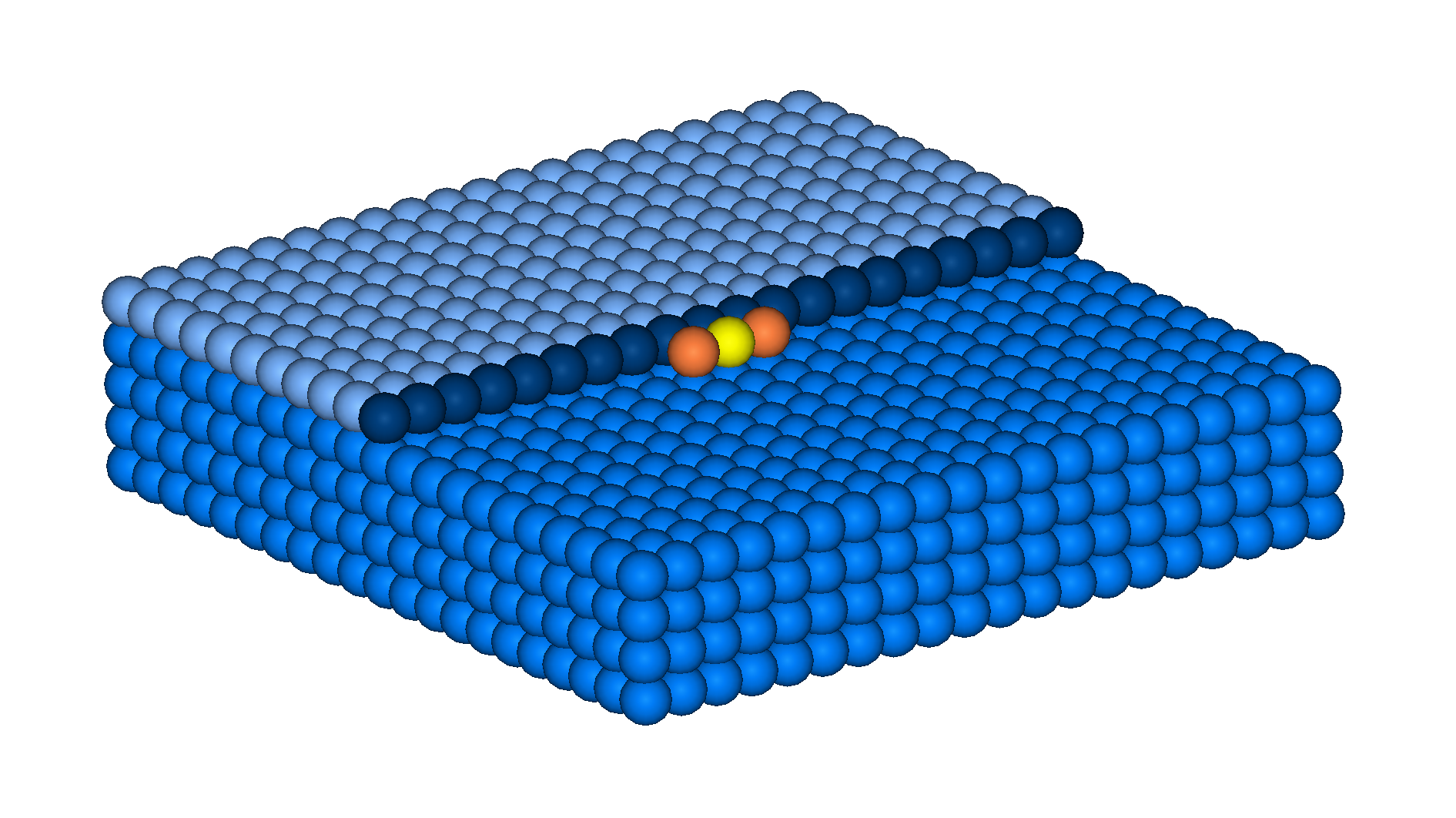}
d)\includegraphics[width=0.45\textwidth]{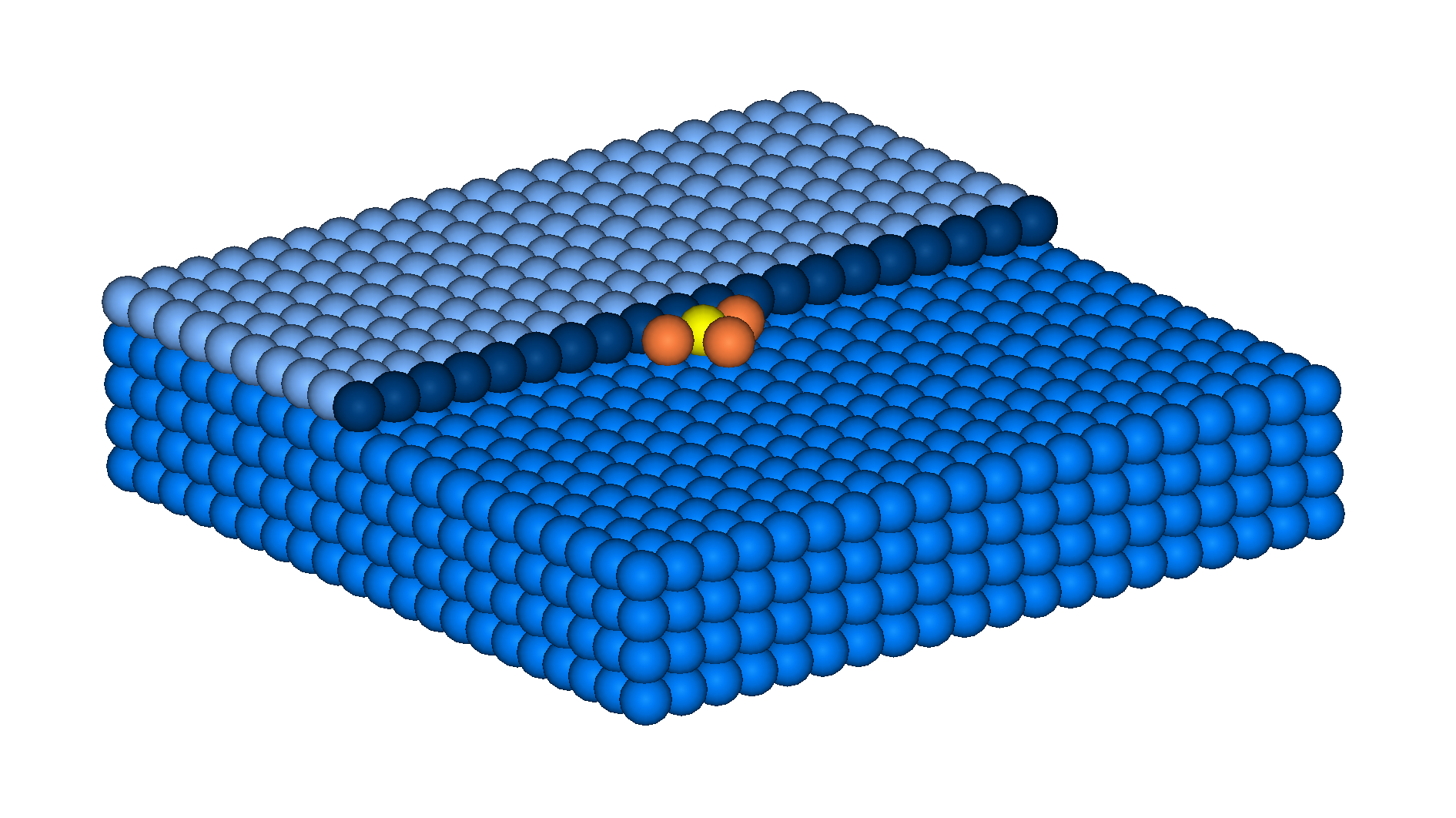}
\caption{Schematic illustration of the selected cellular automaton rules governing the attachment of an adatom (yellow ball) to a crystal (blue balls) in case when a) adatom is situated in a kink in the presence of two additional particles (orange balls); b) adatom has two neighboring crystal sites that do not form a kink in the presence of one additional particle; when adatom has only one neighboring site that forms a crystal in the presence of c) two and d) three additional particles.}
\label{rules}
\end{figure}

We begin the analysis by deciding on the appropriate rules of attachment of adatoms into the structure. We can distinguish two main types - attachment to the kink and attachment to  step. The two main types are further subdivided into different attachment scenarios (CA rules). As previously stated, we perform simulations on a square lattice with a von Neumann neighborhood (sometimes referred to "five-neighbor square"). This approach allows for the consideration of only four nearest neighbors. This results in the following scenarios for adatom attachment: i) The adatom (represented by the yellow ball in Figure~\ref{rules}) is situated in a kink; ii) The adatom has two neighboring crystal sites that do not form a kink (similar to the situation in Figure~\ref{rules}b)); iii) The adatom is surrounded by three neighboring sites that form a crystal; iv) The adatom has only one neighboring crystal site (this case represents a straight step). All of these cases are considered in the absence of any additional particles, as well as in the presence of one, two or three additional particles (orange balls in Figure~\ref{rules}).  In addition, the rules differentiate between the cases where the crystal sites are located on both sides of an adatom or form a kink. Some of these cases are illustrated schematically in Figure~\ref{rules}. In order to ensure clarity and simplicity in the analysis, the process of nucleation of islands on the terrace was disabled. The growth was initiated from seeds located centrally.

All the examples below illustrate scenarios where attachment to the crystal at a kink position occurs immediately - meaning it takes place after completing $n_{DS}$ diffusion jumps, regardless of the presence or absence of neighboring atoms. Kink positions play a crucial role in the crystal growth process.

Among all other CA rules, we have also examined various attachment-to-step events and categorized them into two main classes based on their impact on cluster formation. The first group allows particles to attach anywhere along the step without additional constraints. This class of CA rules leads to the formation of diverse fractal-like structures, exhibiting branching and fine details, which can be more or less densely packed.
On the other hand, if we impose a condition that a single adatom cannot attach to the step unless accompanied by another adatom at a neighboring site, the growth behavior changes. In this case, the resulting structures depend on the local potential, which governs the adatom density. This condition leads to the formation of compact islands with different shapes, determined by both density and local step potential. We will present several examples of such islands.

\subsection{Fractals}
Fractal-like structures emerge when attachment to the step is unconditional - any adatom that reaches the bottom of the step is incorporated into the island.
As a result, the nucleation site at the center of the system expands into  ramified patterns.
Interestingly, even though attachment occurs without any density-dependent constraints, the overall shape of the growing structure is strongly influenced by the depth of the potential well $ E_V $ surrounding it. However, it is not just the depth that plays a role. We will see that the densest structures arise when the local potential outlining the fractal shape is higher, rather than lower, compared to the potential along the surface. We denote the height of this potential by $E_B$. The dependence of the fractal shape on $E_V$ and $E_B$ is somewhat counterintuitive. The higher the local potential, the denser the resulting structure, yet its fractal character remains preserved.

\begin{figure}[hbt]
 \centering
a)\includegraphics[width=0.45\textwidth]{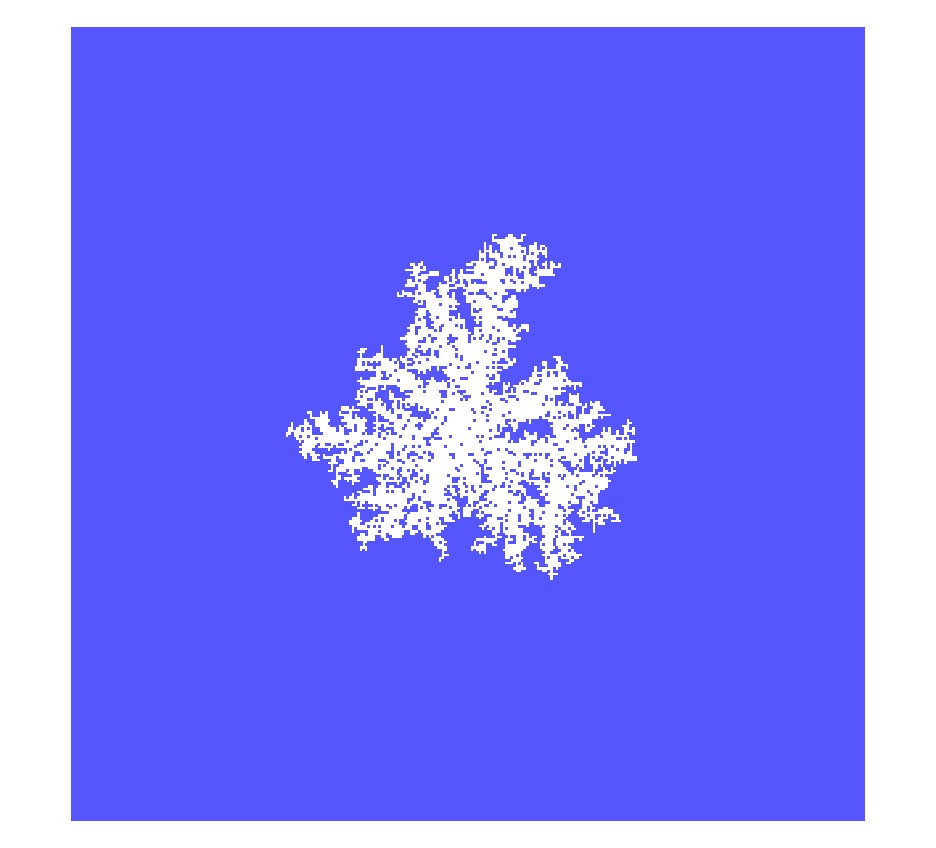}
b)\includegraphics[width=0.45\textwidth]{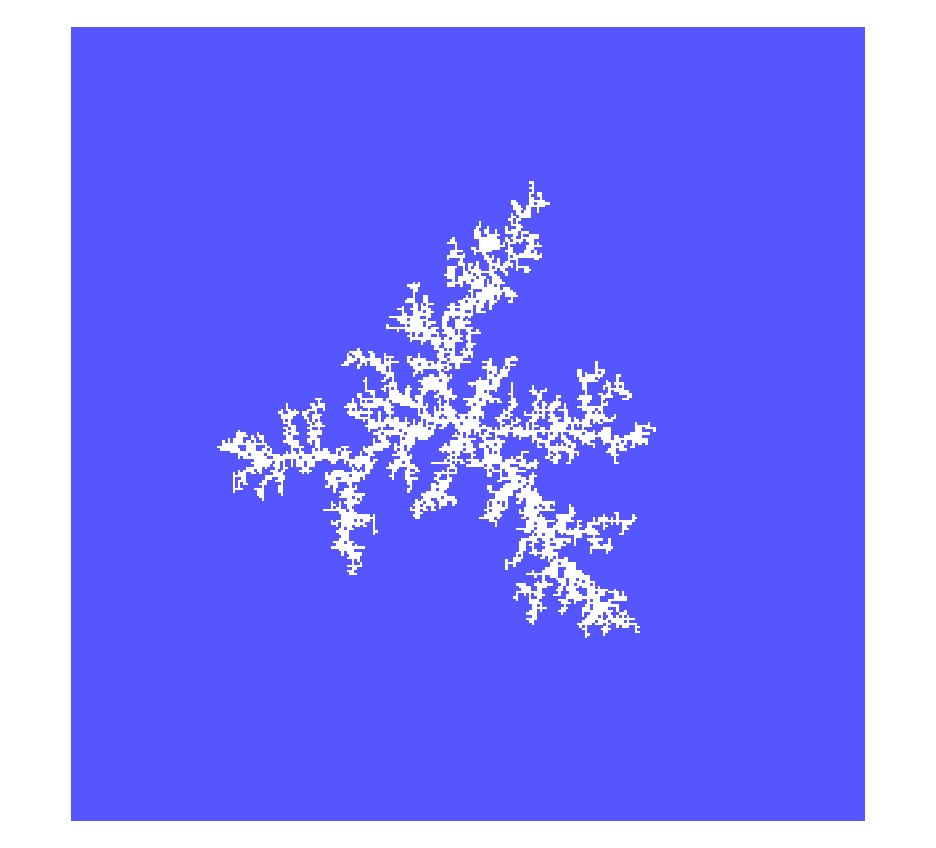}
c)\includegraphics[width=0.45\textwidth]{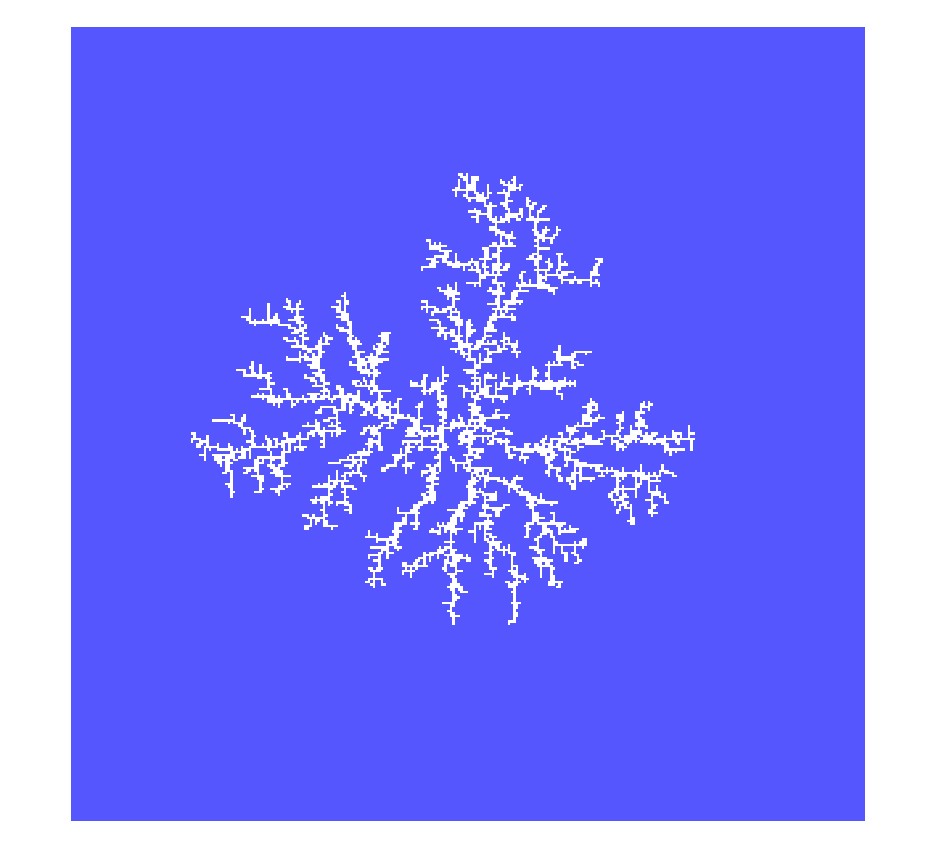}
d)\includegraphics[width=0.45\textwidth]{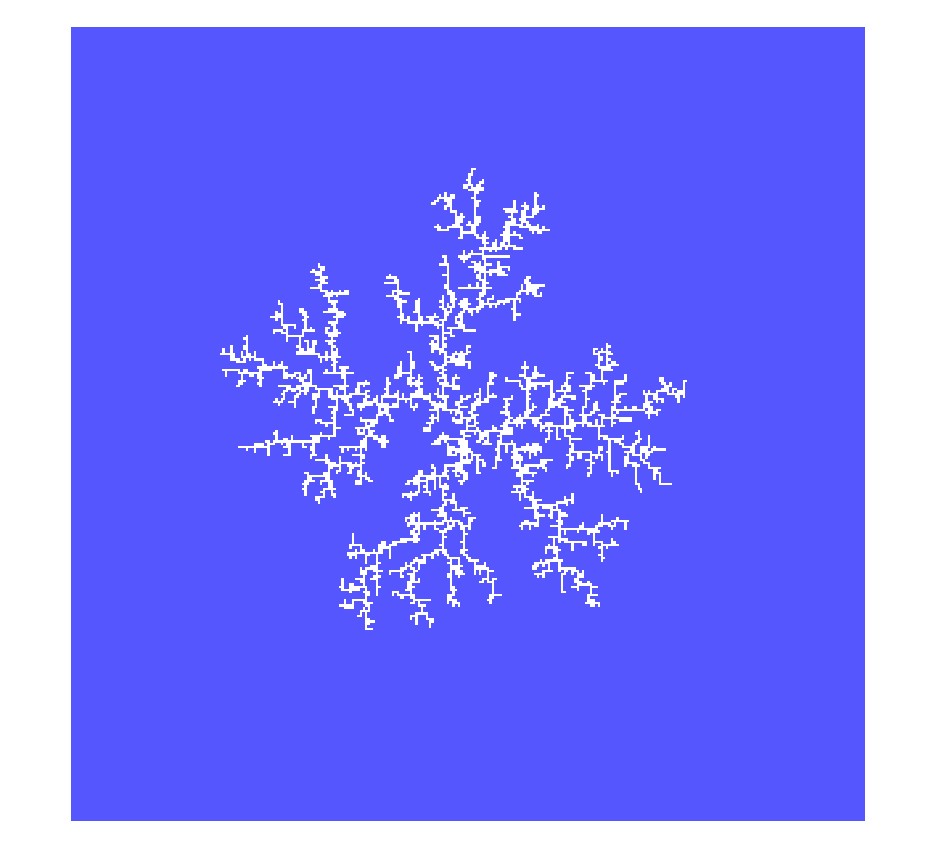}
\caption{The fractal shaped surface patterns obtained for $c_0 = 0.005$, $n_{DS} = 5$, a) $\beta E_B = 2$, b) $\beta E_V=\beta E_B=0$, c) $\beta E_V = 2$, d) $\beta E_V=5$ and a) $2\cdot10^5$, b) $1\cdot 10^5$, c), d) $5\cdot 10^4$ simulation time steps. System size $300 \times 300$.}
\label{fractals}
\end{figure}

The simulations were conducted on a system consisting of $300 \times 300$ lattice sites over $10^5$ simulation time steps. All resulting structures exhibited similar shapes and forms.
 The effects of different potential values  at  the  bottom  of  steps are illustrated in Figure~\ref{fractals}.
All the resulting structures exhibit fractal geometry. As shown in Figure~\ref{fractals}, a deeper potential well leads to a less dense final structure. This can be attributed to the fact that a deeper well reduces the probability of adatom jumps, limiting their mobility. Moreover, when an adatom reaches a position near the step, it is incorporated immediately without any additional conditions, further accelerating the growth process. Note that a lower value of the parameter $\beta E_V$ can result from either a shallower potential $E_V$ or a higher temperature. This implies that, for the same system governed by identical interactions, an increase in temperature leads to the formation of denser fractal structures. These findings are consistent with previously reported results in the literature\cite{Bales,DeVita, Michely}.
Of course, the potential $E_B$, which is higher than the background, characterizes a different class of systems.

\begin{figure}[hbt]
 \centering
\includegraphics[width=0.75\textwidth]{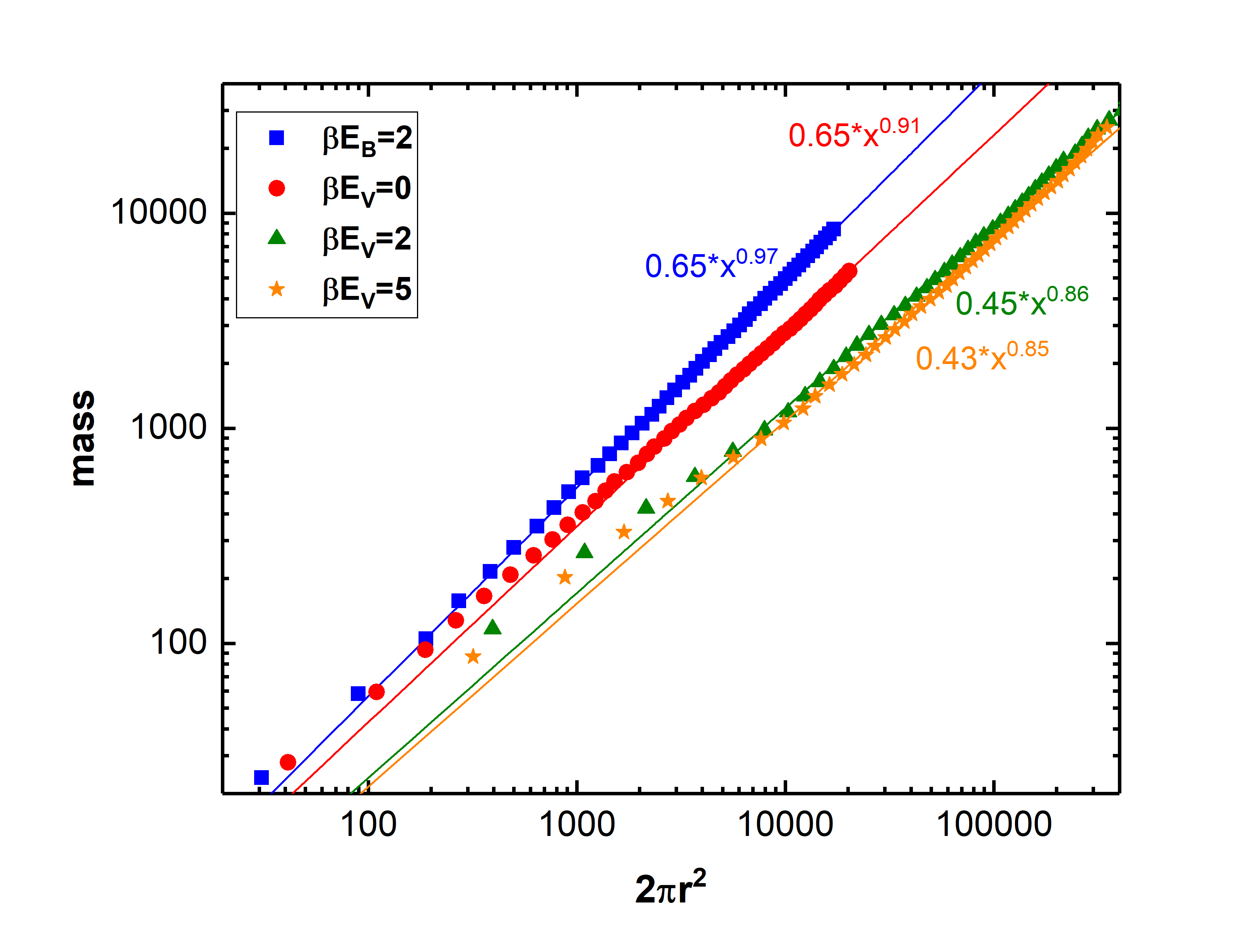}
\caption{The dependence of the number of attached particles on the volume occupied by them for different potential well depths. The results are obtained for $c_0 = 0.005$, $n_{DS} = 5$, $5*10^5$ and $3*10^6$ simulation time steps and averaged over 10 runs.}
\label{fractals-scale}
\end{figure}
It is evident from Figure~\ref{fractals} that fractals grown under different potential depths exhibit variations in structural compactness. While the patterns differ in density, none of them completely fill the space. This raises the question of the dimensionality of such structures. To investigate this, Figure~\ref{fractals-scale} presents the mass of the grown islands - measured as the number of atoms in the structure - plotted against $2\Pi r^2$, where $r$ represents the mean radius over which the structure spreads. The results are averaged over 10 runs. For a fully filled domain, the plots should be linear with an exponent of 1 and a prefactor of 1. If the structure fills space but remains porous, the exponent should still be 1, with the prefactor corresponding to the mean density.

Figure~\ref{fractals-scale} shows four different dependencies for $E_B=2$ and $E_V=0,2,5$. The exponent varies from 0.97 to 0.85, with the highest value observed for the most compact structure at $E_B=2$. This exponent, while close to 1, does not reach it exactly. The prefactor of 0.65 suggests a density in a slightly deformed space characterized by an exponent of 0.97. For $E_V=0$, where no special sites exist at the island boundary, the exponent slightly decreases to 0.91 while maintaining the same prefactor, corresponding to a less compact structure.

Additionally, two fragile patterns emerge for low potentials, $E_V=2$ and $E_V=5$, with exponents of 0.86 and 0.85, respectively, and a prefactors  0.45 and  0.43. These findings suggest that the dimensionality of fractal structures decreases as the potential depth at the bottom of the step increases but eventually saturates for large potential well depths.

\subsection{Compact structures}
\begin{figure}[hbt]
 \centering
a)\includegraphics[width=0.35\textwidth]{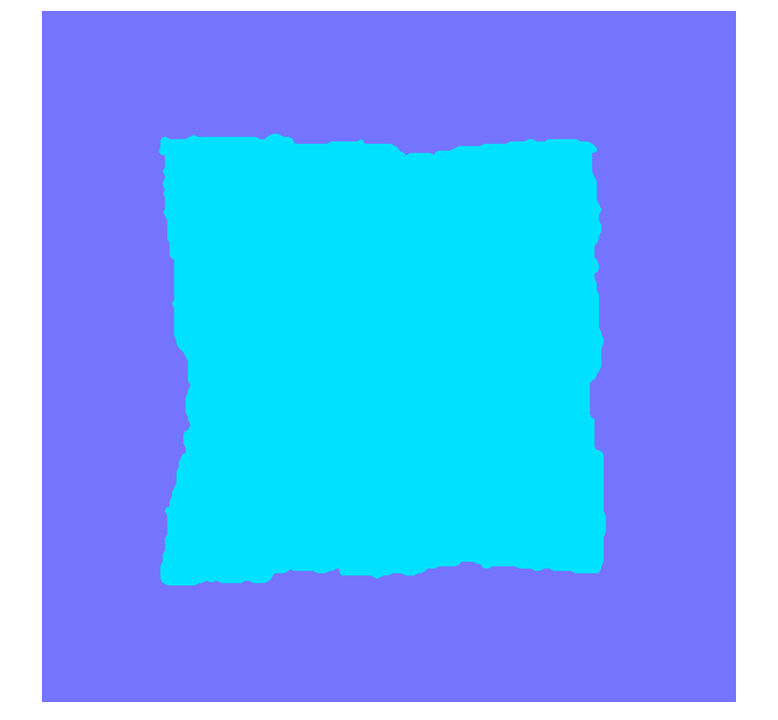}
b)\includegraphics[width=0.35\textwidth]{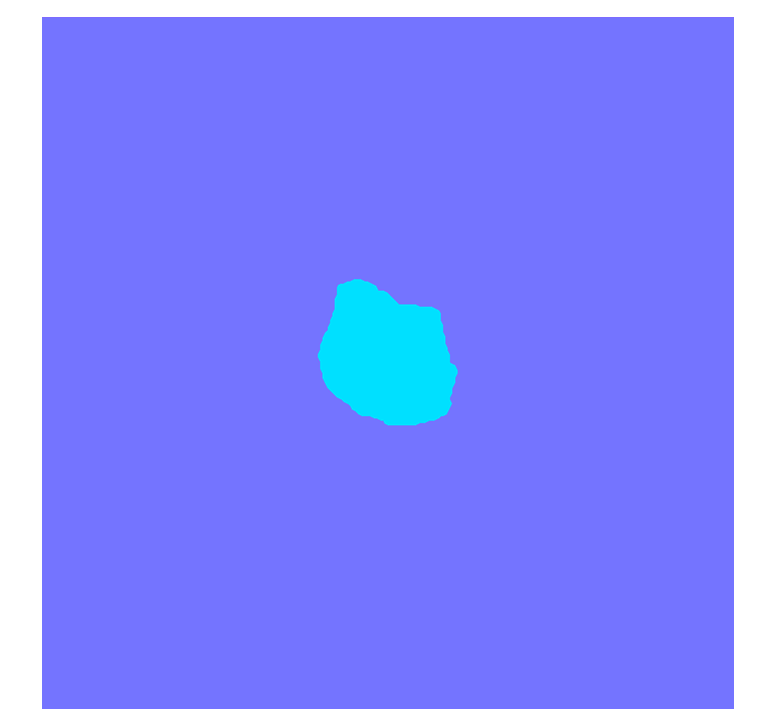}
c)\includegraphics[width=0.35\textwidth]{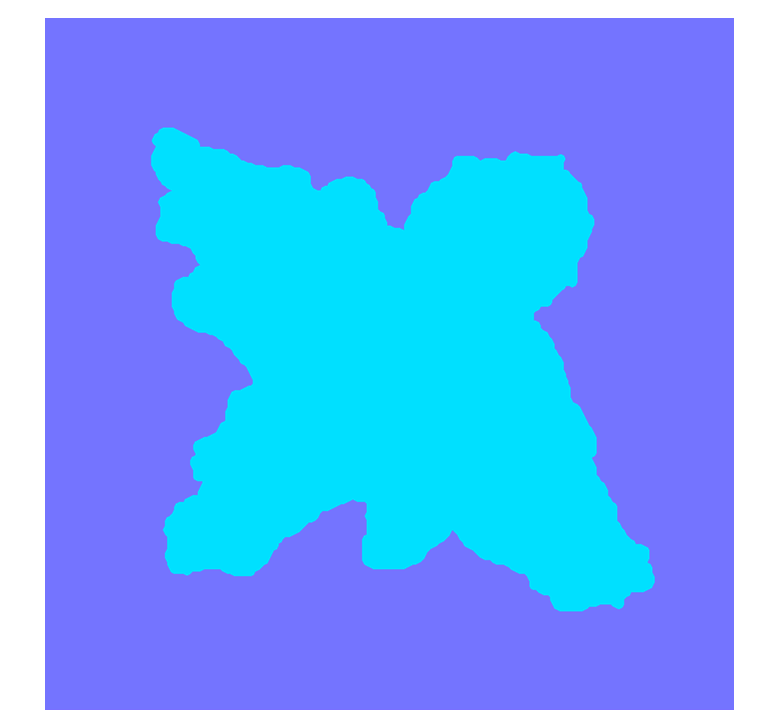}
d)\includegraphics[width=0.35\textwidth]{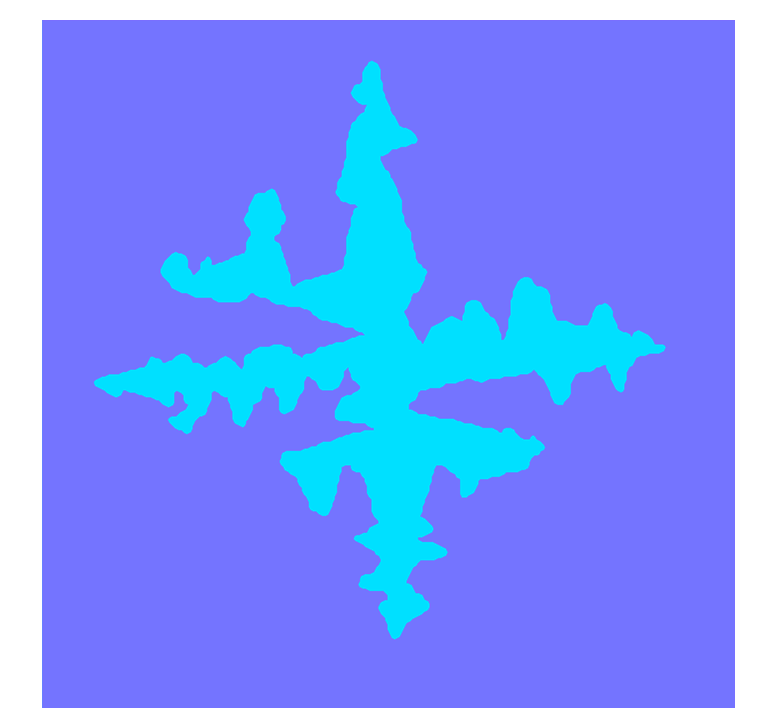}
\caption{The surface patterns obtained for $n_{DS} = 5$, $10^5$ simulation time steps and a) $c_0 = 0.03$, $\beta E_V = 0$, b) $c_0 = 0.005$, $\beta E_V = 4$, c) $c_0 = 0.007$, $\beta E_V = 3$, d) $c_0 = 0.009$, $\beta E_V = 7$. System size $300\times 300$.}
\label{compact}
\end{figure}

In the next stage of our analysis, we examine the attachment of adatoms to a straight step. For a more detailed investigation, we consider the simplest possible attachment rule: an adatom attaches to the step only when accompanied by another adatom at a neighboring site. Additionally, we limit the possible allowed configurations to those where the neighboring atoms are from the left or right side of the attached adatom.
Calculations were performed for different values of the potential at the bottom of the step and various adatom densities, $c_0$. The resulting structures are presented in Figure~\ref{compact}, in which the light blue color is used to denote atoms in a one layer higher.
Firstly, it was observed that modifications in the CA rules led to significant changes in the shape of the islands. The density-dependent attachment of atoms to the steps - implemented as the requirement for a neighboring adatom - promoted the formation of compact structures. Moreover, the resulting patterns exhibited considerable morphological variation depending on the value of $\beta E_V$.
For shallow potential wells at the bottom of the step, square-shaped islands were observed, as shown in Figure~\ref{compact}a. Another possible island shape is rounded, as seen in Figure~\ref{compact}b, which emerges when the potential well is slightly deeper, consistent with previous findings in Ref.\cite{Albe}. A further increase in potential well depth results in elongated island shapes aligned with the crystal lattice directions (see Figure~\ref{compact}c). These structures exhibit a square-like geometry with sides compressed toward the center, which, in three dimensions, are commonly referred to as Hooper shapes \cite{Yang-AFM}. At even higher values of
$\beta E_V$, islands evolve into dendritic structures, consistent with previous reports in the literature \cite{Bales,Michely,Yang-AFM,Albe}.

\begin{figure}[hbt]
 \centering
\includegraphics[width=0.5\textwidth]{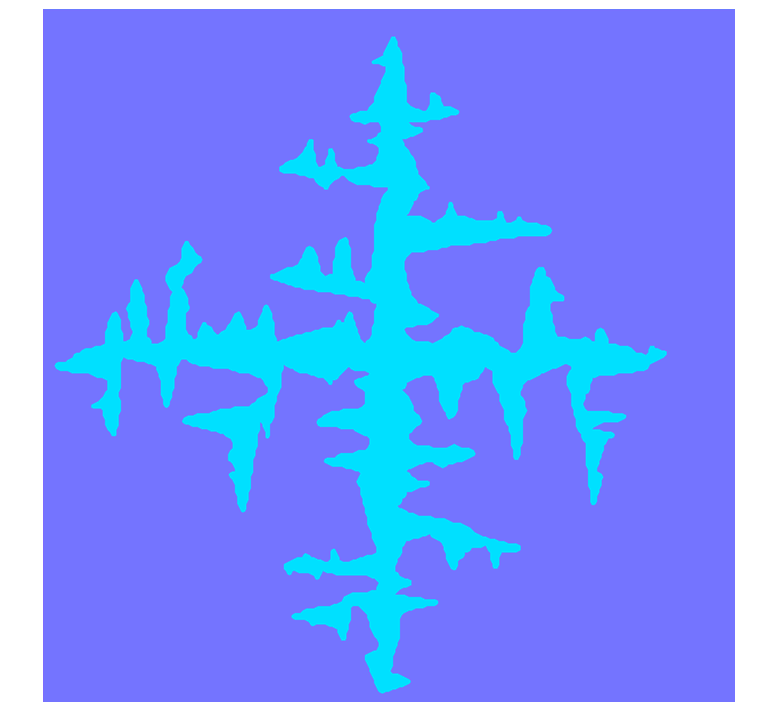}
\caption{The dendritic shape island on a crystal surface obtained in the case of attachment of adatom when neighboring sites are filled by one particles, independent on the particle position for $n_{DS} = 5$, $c_0 = 0.007$, $\beta E_V = 11$, $10^5$ simulation time steps. System size $300\times 300$.}
\label{dendrite-dense}
\end{figure}

If we focus only on dendritic structures and allow adatoms to attach the step independently on the position of an additional particle, so  additional  adatom can be  in one  out  of  four  possible positions, we observed a change in the final morphology. The dendritic structures are much more ramified (see Figure~\ref{dendrite-dense}) than in the previous case.

\begin{figure}[hbt]
 \centering
\includegraphics[width=0.85\textwidth]{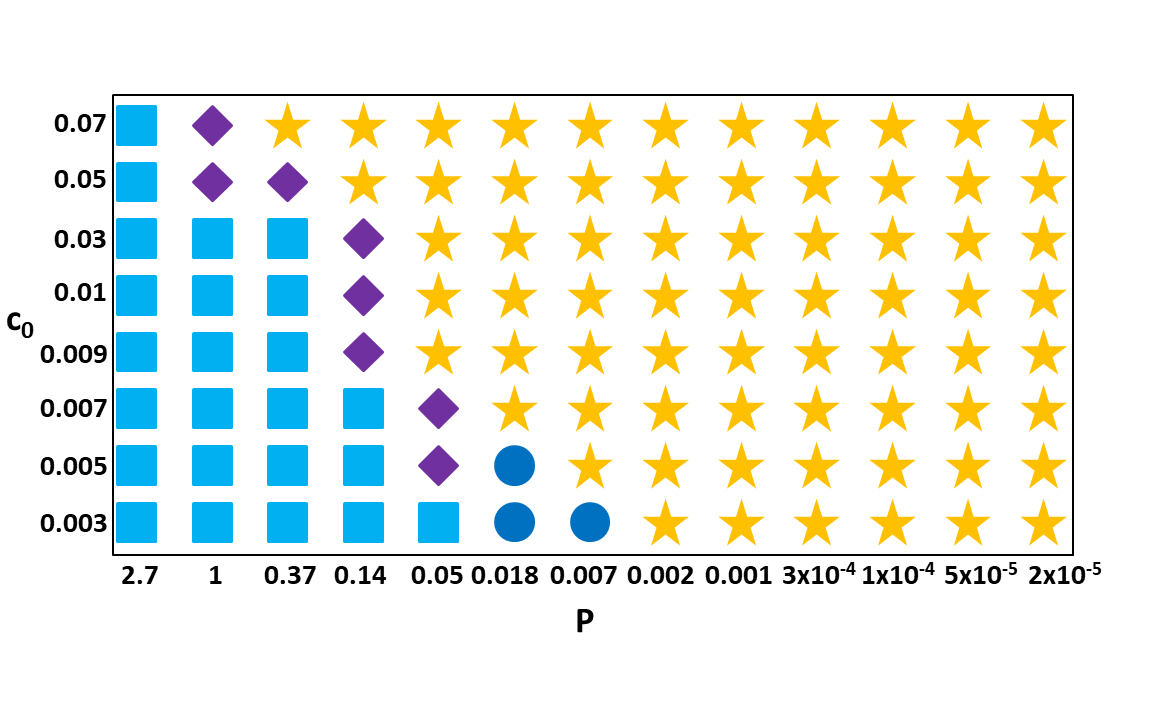}
\caption{Diagram of the formation of various islands morphologies obtained for $n_{DS} = 5$ and $t = 10^5$. Different symbols correspond to different patterns:  {\color{cyan}$\blacksquare$} describes rectangular shape, {\color{violet}$\blacklozenge$} describe Hopper-shape structures, {\color{NavyBlue}$\CIRCLE$} means rounded structures and {\color{YellowOrange}$\bigstar$} means dendritic structures.}
\label{diagram}
\end{figure}

A summary of the obtained results, as  a function  of initial particle concentrations and the probability of jump out the well P is presented in the Figure~\ref{diagram}, which shows a diagram of the various island morphologies observed. We recall that $P=e^{-\beta E_V}$ or in the case of the potential higher that the background $P=e^{\beta E_B}$.
As shown in the diagram, two primary regions can be identified, characterized by the formation of square- or dendritic-shaped islands (shown in Figure~\ref{compact}a and d). Between them, two other island types can be distinguished: round and Hooper-shaped structures, shown in Figure~\ref{compact}b and c.

Square-shaped islands are associated with shallow energy potential wells,
while more rounded domain shapes emerge when the potential is deeper. A deeper potential well accumulates more adatoms near the island boundary, suggesting that its effect can be approximated by an increase in adatom density. Indeed, this tendency is visible in the diagram in Figure~\ref{diagram}.

It can be observed that for higher initial adatom concentrations, the energy potential well must be shallower to maintain the same island morphology. Dendritic islands are obtained for deep energy potential wells, which can be interpreted as corresponding to lower temperatures or higher adatom densities. Conversely, square or rectangular islands form in systems with shallow potential wells or even a low potential hill at the boundary and at low adatom densities. In intermediate parameter regions, islands with rounded or Hooper-shaped morphologies are observed.

\subsection{Growth with nucleation}
In the previous examples, individual domains of various shapes were analyzed to investigate their geometry and the parameters responsible for the formation of specific structures. To focus on the shape and growth dynamics of a single domain, spontaneous nucleation was excluded.
\begin{figure}[hbt]
 \centering
a)\includegraphics[width=0.35\textwidth]{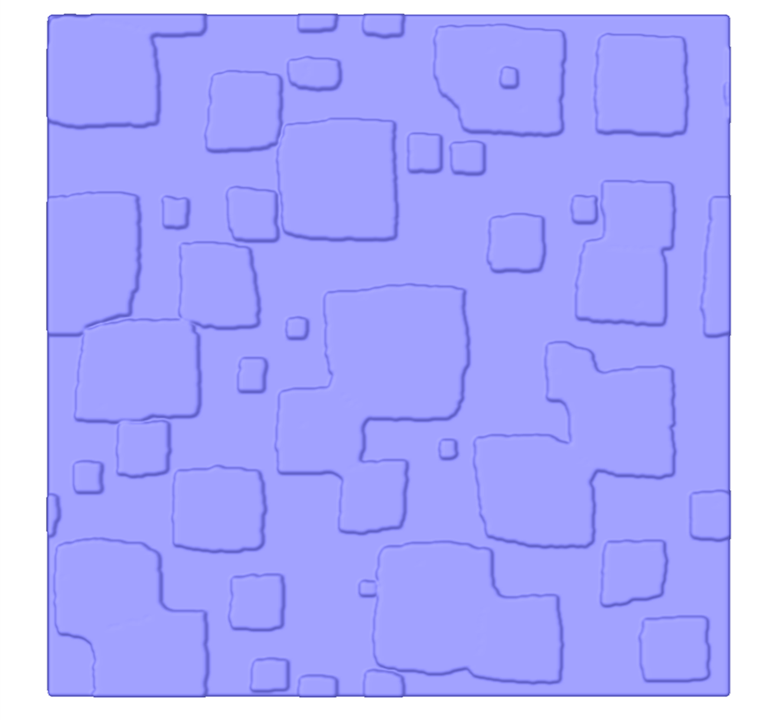}
b)\includegraphics[width=0.35\textwidth]{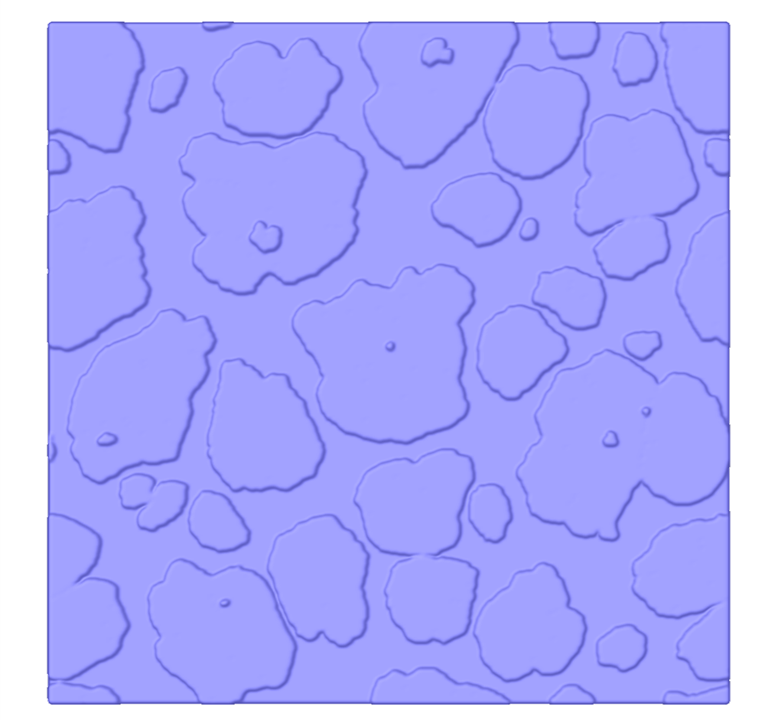}
c)\includegraphics[width=0.35\textwidth]{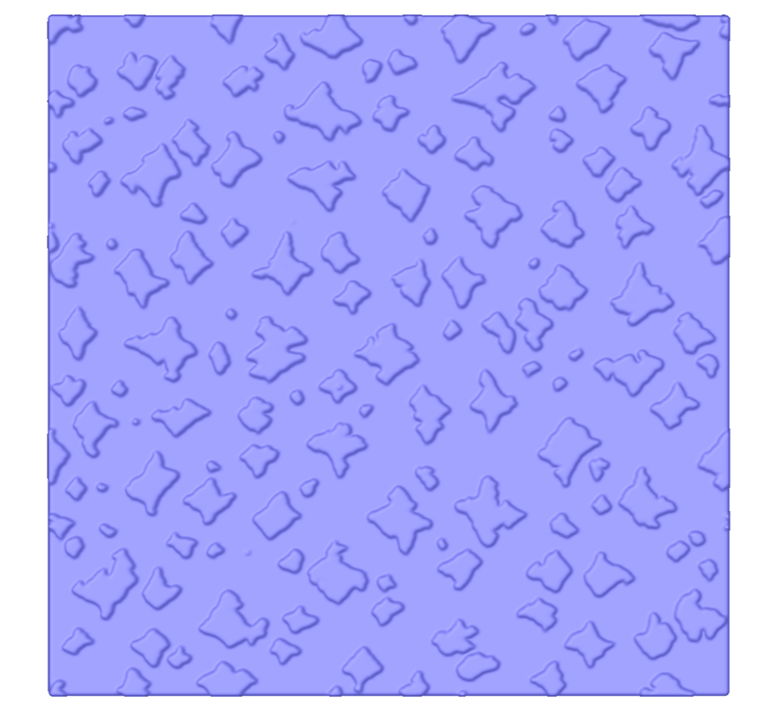}
d)\includegraphics[width=0.35\textwidth]{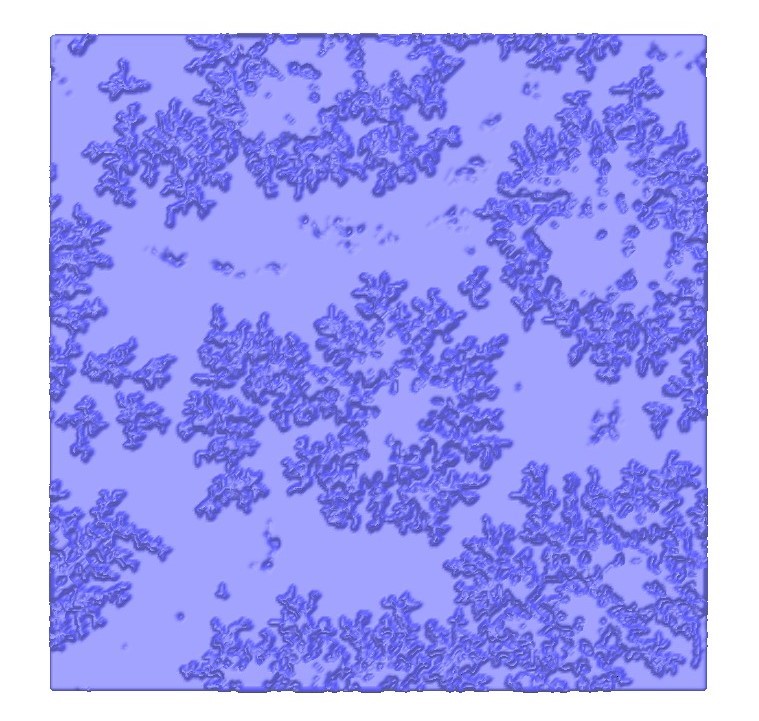}
\caption{The surface patterns on crystal surface obtained in case of allowed nucleation of islands on a terrace for $n_{DS} = 5$, a) $c_0 = 0.003$, $\beta E_V = 1$ and t=$4.1\cdot10^5$; b) $c_0 = 0.003$, $\beta E_V = 4$ and t=$2.6\cdot10^5$; c)  $c_0 = 0.009$, $\beta E_V = 6$, and t=$1.4\cdot10^4$; d)  $c_0 = 0.03$, $\beta E_V = 4$, and t=$2\cdot 10^5$. System size $300\times 300$. }
\label{nucleated}
\end{figure}

Now, we examine what happens when domains nucleate across the surface and form successive layers on top of the initial one. Three panels in Figure~\ref{nucleated}, a, b and c illustrate the early stages of growth for square, round and dendritic structures. We observe numerous small islands with characteristic shapes, along with newly formed ones stacking atop them.

A fourth example, shown in Figure~\ref{nucleated}d, presents a surface covered by fractal-like structures after prolonged evolution. It is evident that as additional crystal layers gradually fill with atoms, the delicate fractal-like pattern remains on the top as a convex structure, while a corresponding concave hole pattern forms at the bottom.

\section{Conclusions}

The vicCA model was used to analyze the spontaneous formation of islands on a surface, with growth dynamics governed by adatom attachment rules. Two primary attachment modes play a crucial role in this process: attachment at kinks and attachment at steps. Adatoms preferentially attach at kink positions, making these sites key determinants of crystal growth behavior.

When attachment to steps is unrestricted, fractal-like structures emerge as adatoms incorporate into the growing island without density constraints. The density of these fractals depends on the potential well depth $E_V$ surrounding the structure. Higher local potential $E_B$ results in denser formations, though the overall fractal nature persists. Increasing temperature (lower $\beta E_V$) further enhances density, consistent with previous studies.

Introducing a constraint that requires an adatom to be accompanied by another at a neighboring site alters the growth pattern. This condition promotes compact island formation, with morphology shaped by $\beta E_V$.
Different potential well depths yield distinct island shapes, with shallow wells forming square islands, moderate depths producing rounded or Hooper-shaped islands, and deep wells leading to dendritic growth.
Spontaneous nucleation leads to the formation of multiple islands across the surface. Over time, these islands stack, forming layered structures with varying surface roughness. A systematic study of island formation as a function of initial adatom concentration and potential well depth  reveals distinct morphological regions. Higher adatom concentrations require shallower potential wells to maintain similar growth patterns.
The transition between square, rounded, Hooper, and dendritic structures aligns with variations in $\beta E_V$ and initial adatom concentration.

This study confirms that crystal growth patterns are determined by the interplay between attachment rules, adatom density, and local potential depth. The findings align with previous research, demonstrating that nucleation and step attachment mechanisms significantly shape final surface structures. Future research could refine the model to explore additional factors, such as diffusion constants, temperature gradients, or external field effects.

\section*{Acknowledgments}
The authors express their gratitude to The Polish National Center for Research and Development (grant no. EIG CONCERT-JAPAN/9/56/AtLv-AlGaN/2023) for providing financial support.
They also extend their thanks to Vesselin Tonchev from the Faculty of Physics at Sofia University, Hristina Popova from Institute of Physical Chemistry of Bulgarian Academy of Sciences, Anna Nied\'{z}wiecka from Institute of Physics, Polish Academy of Sciences and Yoshihiro Kangawa from the Research Institute for Applied Mechanics at Kyushu University for their valuable discussions.


\begin{thebibliography}{00}
\bibitem{Chen} M. Chen, T. Z. Kattamis, Dendrite coarsening during directional solidification of Al-Cu-Mn alloys, Mater. Sci. Eng. A247 (1998) 239-247. https://doi.org/10.1016/S0921-5093(97)00720-X.
\bibitem{Mendoza} R. Mendoza, J. Alkemper, P. W. Voorhees, The morphological evolution of dendritic microstructures during coarsening, Metal. Mater. Trans. 34 (2003) 481-489. https://doi.org/10.1007/s11661-003-0084-2.
\bibitem{Kammer} D.Kammer, and P.W. Voorhees, The morphological evolution of dendritic microstructures during coarsening, Acta Mater. 54 (2006) 1549-1558. https://doi:10.1016/j.actamat.2005.11.031
\bibitem{Zhang} H. Zhang, Q. Xu, Multi-scale simulation of directional dendrites growth in superalloys, J. Mater. Proces. Technol. 238 (2016) 132-141. http://dx.doi.org/10.1016/j.jmatprotec.2016.07.013.
\bibitem{Xiao-88} R.-F. Xiao, J. I. D. Alexander, F. Rosenberger, Morphological evolution of growing crystals: A Monte Carlo simulation, Phys. Rev. A 100 (1988) 313. https://doi.org/10.1103/PhysRevA.38.2447.
\bibitem{Bales} G. S. Bales, D. C. Chrzan, Transition from compact to fractal islands during submonolayer epitaxial growth, Phys. Rev. Lett 74 (1995) 4879. https://doi.org/10.1103/PhysRevLett.74.4879.
\bibitem{DeVita} J. P. DeVita, L. M. Sander, P. Smereka, Quasicontinuum Monte Carlo simulation of multilayer surface growth, in: A. Voigt (Ed.), Multiscale Modeling in Epitaxial Growth, Birkh\"auser Verlag, Basel - Boston - Berlin, 2000, pp 57-66.
\bibitem{Michely}  T. Michely, J. Krug, Islands, mounds and atoms, Springer, Berlin, 2004.
\bibitem{Wang} K. Wang, G. Pr\'evot, J.-N. Aqua, Anomalous intralayer growth of epitaxial Si on Ag(111), Sci. Rep. 14 (2014) 2401.
\bibitem{Yang-AFM}  Z. Yang, J. Zhang, L. Zhang, B. Fu, P. Tao, Ch. Song, W. Shang, T. Deng, Self-assembly in Hopper-shaped crystals, Adv. Funct. Mater. 30 (2020) 1908108. https://doi.org/10.1002/adfm.201908108.
\bibitem{Klepka} B. P. Klepka, A. Micha\'s, T. Wojciechowski, A. Nied\'zwiecka, Extremely charged coral protein AGARP regulates calcium carbonate growth through liquid phase separation, bioRxiv 2024.06.04.597398. https://doi.org/10.1101/2024.06.04.597398.
\bibitem{Akiyama-JJAP} T. Akiyama, T. Kawamura, Ab initio study for adsorption behavior on AlN(0001) surface with steps and kinks during metal-organic vapor-phase epitaxy, Jpn. J. Appl. Phys. 63 (2024) 02SP71. https://doi.org/10.35848/1347-4065/ad1896.
\bibitem{Krasteva-AIP} A. Krasteva, H. Popova, F. Krzy\.{z}ewski, M. Za{\l}uska-Kotur, V. Tonchev, Unstable vicinal crystal growth from cellular automata, AIP Conf. Proc. 1722 (2016) 220014. https://doi.org/10.1063/1.4944246.
\bibitem{Krzyzew-JCG} F. Krzy\.{z}ewski, M. Za{\l}uska-Kotur, A. Krasteva, H. Popova, V. Tonchev, Step bunching and macrostep formation in 1D atomistic scale model of unstable vicinal crystal growth, J. Cryst. Growth 474 (2017) 135-139. https://doi.org/10.1016/j.jcrysgro.2016.11.121.
\bibitem{Toktarbaiuly-prb} O. Toktarbaiuly, V. Usov, C. Coile\'{a}in, K. Siewierska, S. Krasnikov, E. Norton, S. I. Bozhko, V. N. Semenov, A. N. Chaika, B. E. Murphy, O. L\"{u}bben, F. Krzy\.{z}ewski, M. A. Za{\l}uska-Kotur, A. Krasteva, H. Popova, V. Tonchev, I. V. Shvets, Step bunching with both directions of the current: Vicinal W(110) surfaces versus atomistic-scale model, Phys. Rev. B 97 (2018) 035436.  https://doi.org/10.1103/PhysRevB.97.035436.
\bibitem{Krzyzew-CGD} F. Krzy\.{z}ewski, M. Za{\l}uska-Kotur, A. Krasteva, H. Popova, V. Tonchev, Scaling and dynamic stability of model vicinal surfaces, Cryst. Growth Des. 19 (2019) 821-831. https://doi.org/10.1021/acs.cgd.8b01379.
\bibitem{Popova-CGD} H. Popova, F. Krzy\.{z}ewski, M. A. Za{\l}uska-Kotur, V. Tonchev, Quantifying the effect of step-step exclusion on dynamically unstable vicinal surfaces: step bunching without macrostep formation, Cryst. Growth Des. 20 (2020) 7246-7259. https://doi.org/10.1021/acs.cgd.0c00927.
\bibitem{Popova-CGD23} H. Popova, Analyzing the pattern formation on vicinal surfaces in diffusion-limited and kinetics-limited growth regimes: The effect of step-step exclusion, Cryst. Growth Des. 23 (2023) 8875-8888. https://doi.org/10.1021/acs.cgd.3c00952.
\bibitem{MZK-crystals} M. Za{\l}uska-Kotur, H. Popova, V. Tonchev, Step Bunches, Nanowires and other vicinal "creatures" -- Ehrlich--Schwoebel effect by cellular automata, Crystals 11 (2021) 1135. https://doi.org/10.3390/cryst11091135.
\bibitem{Chabowska-ACS} M. A. Chabowska, M. A. Za{\l}uska-Kotur, Diffusion-dependent pattern formation on crystal surfaces, ACS Omega 8 (2023) 45779-45786. https://doi.org/10.1021/acsomega.3c06377.
\bibitem{Chabowska-Vac} M. A. Chabowska, M. A. Za{\l}uska-Kotur, Surface patterns shaped by additives in crystals, arXiv:2409.04084. https://doi.org/10.48550/arXiv.2409.04084.
\bibitem{Chabowska-PRB} M. A. Chabowska, H. Popova, M. A. Za{\l}uska-Kotur, Step meandering: the balance between the potential well and the Ehrlich -- Schwoebel barrier, arXiv:2411.12487. https://doi.org/10.48550/arXiv.2411.12487.
\bibitem{Ohka-CGD} T. Ohka, T. Akiyama, A. Muizz Pradipto, K. Nakamura, T. Ito, Effect of step edges on adsorption behavior for GaN(0001) surfaces during metalorganic vapor phase epitaxy: An ab initio study, Cryst. Growth Des. 20 (2020) 4358-4365. https://doi.org/10.1021/acs.cgd.0c00117.
\bibitem{Akiyama-JCG20} T. Akiyama, T. Ohka, K. Nakamura, T. Ito, Ab initio study for adsorption and desorption behavior at step edges of GaN (0001) surface, J. Cryst. Growth 532 (2020) 125410. https://doi.org/10.1016/j.jcrysgro.2019.125410.
\bibitem{Albe} K. Albe, M. M\" uller, Cluster diffusion and island formation on fcc(111) metal surfaces studied by atomic scale computer simulations, in: A. Voigt (Ed.), Multiscale Modeling in Epitaxial Growth, Birkh\"auser Verlag, Basel - Boston - Berlin, 2000, pp 19-28.

\end{thebibliography}
\end{document}